\documentclass[12pt]{article}%
\usepackage{amsmath}%
\usepackage{amsfonts}%
\usepackage{amssymb}%
\usepackage{graphicx}
\newtheorem{theorem}{Theorem}
\newtheorem{acknowledgement}[theorem]{Acknowledgement}

\newtheorem{corollary}[theorem]{Corollary}

\newtheorem{definition}[theorem]{Definition}

\newtheorem{lemma}[theorem]{Lemma}

\newtheorem{proposition}[theorem]{Proposition}
\newtheorem{remark}[theorem]{Remark}

\newenvironment{proof}[1][Proof]{\textbf{#1.} }{\ \rule{0.5em}{0.5em}}

\begin{document}

\title{Separation of variables in quasi-potential systems of bi-cofactor form}
\author{Krzysztof Marciniak\thanks{On leave of absence from Department of Physics, A.
Mickiewicz University, Pozna\'{n}, Poland.}\\Department of Science and Technology\\Campus Norrk\"{o}ping, Link\"{o}ping University\\601 74 Norrk\"{o}ping, Sweden\\krzma@itn.liu.se
\and Maciej B\l aszak\thanks{Partially supported by KBN grant No 5P03B 004 20}\\Department of Physics\\A. Mickiewicz University\\Umultowska 85, 61-614 Pozna\'{n}\\blaszakm@main.amu.edu.pl}
\maketitle

\begin{abstract}
We perform variable separation in the quasi-potential systems of equations of
the form $\ddot{q}=-A^{-1}\nabla k=-\tilde{A}^{-1}\nabla\tilde{k}${} , where
$A$ and $\tilde{A}$ are Killing tensors, by embedding these systems into a
bi-Hamiltonian chain and by calculating the corresponding Darboux-Nijenhuis
coordinates on the symplectic leaves of one of the Hamiltonian structures of
the system. We also present examples of the corresponding separation
coordinates in two and three dimensions.

\end{abstract}

\bigskip AMS 2000 Classification System: 70H06,70H20,53D17

\section{Introduction}

In recent years, a new constructive separability theory, based on a
bi-Hamiltonian property of integrable systems, was presented. In the frame of
canonical coordinates the theory was developed in a series of papers
\cite{1}-\cite{6} (see also the review article \cite{7}), while a general case
was considered in \cite{m1} - \cite{8}.

In this paper we apply the bi-Hamiltonian separability theory to solve
Hamilton-Jacobi (HJ) equations for a class of quasi-potential systems, called
bi-cofactor systems, systematically studied in recent papers \cite{PoissonJMP}%
-\cite{oscylator}.

The structure of the paper is as follows. In Section 2 we briefly describe -
the one-Casimir bi-Hamiltonian separability theory in case of bi-Hamiltonian
systems with quadratic in momenta constants of motion and relate this theory
with the classical St\"{a}ckel theory. In Section 3 we present basic facts
about quasi-potential systems and their special subclass called bi-cofactor
systems. Then, in Section 4, an explicit form of transformation to separated
coordinates is derived, proving the separability of all nondegenerated
bi-cofactor systems. We sketch the proofs of more important results of this
section independently of the general statements of Section 2 in order to make
this article more self-contained, but we also point out the places where the
general theory and our calculations meet. Finally, in Section 5, we illustrate
the obtained formulas by some examples.

\section{St\"{a}ckel separability of one-Casimir bi-Hamiltonian chains}

Let us consider a Poisson manifold $\mathcal{M}$ of $\dim\mathcal{M}=2n+1$
equipped with a linear Poisson pencil $\Pi_{\xi}$ $=\Pi_{1}-\xi\Pi_{0}$ of
maximal rank, i.e. a pair of Poisson operators (tensors) $\Pi_{i}:T^{\ast
}\mathcal{M}\rightarrow T\mathcal{M}$ each of rank $2n$ such that their linear
combination $\Pi_{1}-\xi\Pi_{0}$ is itself a Poisson operator for any $\xi
\in\mathbf{R}$ (the operators $\Pi_{0}$ and $\Pi_{1} $ are then said to be compatible).

\begin{definition}
\label{casimirdef}A scalar function $h:\mathcal{M}\rightarrow\mathbf{R}$ is
Called a Casimir function (or a Casimir) of a Poisson operator $\Pi$ acting in
$\mathcal{M}$ \ if $\Pi\circ dh=0$.
\end{definition}

Here and in what follows $d:C^{\infty}(\mathcal{M})\rightarrow T^{\ast
}\mathcal{M}$ is the operator of external derivative (gradient operator) in
$\mathcal{M}$ and the symbol $\circ$ denotes the composition of mappings. If -
as it often happens - the Casimir $h_{\xi}$ of the pencil $\Pi_{\xi}$ is a
polynomial in $\xi$ of order $n$
\begin{equation}
h_{\xi}=h_{n}\xi^{n}+h_{n-1}\xi^{n-1}+...+h_{0}\label{2}%
\end{equation}
the by expanding the equation $\Pi_{\xi}\circ dh_{\xi}=0$ in powers of $\xi$
and comparing the coefficients at equal powers we obtain the following
\emph{bi-Hamiltonian chain}:
\begin{equation}%
\begin{array}
[c]{ccc}%
\Pi_{1}\circ dh_{0} & = & 0\\
\Pi_{1}\circ dh_{1} & = & \Pi_{0}\circ dh_{0}\\
& \vdots & \\
\Pi_{1}\circ dh_{n} & = & \Pi_{0}\circ dh_{n-1}\\
0 & = & \Pi_{0}\circ dh_{n}%
\end{array}
\label{3new}%
\end{equation}
From (\ref{3new}) it follows that the functions $h_{i}$ are in involution with
respect to both Poisson structures. If additionally all $h_{i}$ are
functionally independent, then the chain is a Liouville integrable system.

Let us now consider a set of coordinates $\{\lambda_{i},\mu_{i}\}_{i=1}^{n} $
and a Casimir coordinate $c^{\prime}=h_{n}$ on $\mathcal{M}$ so that
$\{\lambda_{i},\mu_{i}\}_{i=1}^{n}$ are canonical with respect to $\Pi_{0}$,
so that%

\[
\Pi_{0}=\left[
\begin{array}
[c]{c|c}%
\begin{array}
[c]{cc}%
0 & I\\
-I & 0
\end{array}
&
\begin{array}
[c]{c}%
0\\
0
\end{array}
\\\hline
\ast & 0
\end{array}
\right]
\]
We can try to linearize the system (\ref{3new}) through a canonical
transformation $(\lambda,\mu)\rightarrow(b,a)$ in the form $b_{i}%
=\frac{\partial W}{\partial a_{i}},\mu_{i}=\frac{\partial W}{\partial
\lambda_{i}},$ where $W(\lambda,a)$ is a generating function that solves the
related Hamilton-Jacobi (HJ) equations
\begin{equation}
h_{r}(\lambda,\frac{\partial W}{\partial\lambda})=a_{r}%
,\;\;\;\;\;r=0,...,n.\label{5}%
\end{equation}
In general, the HJ equations (\ref{5}) are nonlinear partial differential
equations that are very difficult to solve. However, there are rare cases when
one can find \ a solution of (\ref{5}) \ in the separable form
\begin{equation}
W(\lambda,a)=\sum_{i=1}^{n}W_{i}(\lambda_{i},a)\label{6}%
\end{equation}
that turns the HJ equations into a set of decoupled ordinary differential
equations that can be solved by quadratures. Such $(\lambda,\mu)$ coordinates
are called \emph{separated coordinates. }In the $(a,b)$ coordinates the flow
$d/dt_{j}$ associated with every Hamiltonian $h_{j}$ is trivial
\begin{equation}
\frac{da_{i}}{dt_{j}}=0,\text{ \ }\frac{db_{i}}{dt_{j}}=\delta_{ij}\text{ ,
}\frac{dc^{\prime}}{dt_{j}}=0,\text{\ \ }i,j=1,\ldots,n\label{8}%
\end{equation}
and the implicit form of the trajectories $\lambda(t_{j})$ is given by
\begin{equation}
b_{i}(\lambda,a)=\frac{\partial W}{\partial a_{i}}=\delta_{ij}t_{j}%
+const,\,\,\,\,i,j=1,...,n.\label{9}%
\end{equation}

\begin{theorem}
\textbf{\ } \label{iff} A sufficient condition for $(\lambda,\mu)$ to be
separated coordinates for the bi-Hamiltonian chain (\ref{3new}) is
\begin{equation}
H_{i}(\lambda,\mu,c^{\prime})=f_{i}(\lambda_{i},\mu_{i}%
),\;\;\;i=1,...,n,\label{11}%
\end{equation}
where
\[
H_{i}(\lambda,\mu,c^{\prime})\equiv c^{\prime}\lambda_{i}^{n}+h_{n-1}%
(\lambda,\mu,c^{\prime})\lambda_{i}^{n-1}+...+h_{0}(\lambda,\mu,c^{\prime})
\]
and $f_{i}(\lambda_{i},\mu_{i})$ is some smooth function of a pair of
canonically conjugate coordinates $\lambda_{i},\mu_{i}.$
\end{theorem}

We will briefly sketch the proof here. The condition (\ref{11}) can be
presented in a matrix form
\begin{equation}
\left(
\begin{array}
[c]{cccc}%
\lambda_{1}^{n} & \lambda_{1}^{n-1} & \cdots & 1\\
\lambda_{2}^{n} & \lambda_{2}^{n-1} & \cdots & 1\\
\vdots & \vdots & \cdots & \vdots\\
\lambda_{n}^{n} & \lambda_{n}^{n-1} & \cdots & 1
\end{array}
\right)  \left(
\begin{array}
[c]{c}%
c\\
h_{n-1}\\
\vdots\\
h_{0}%
\end{array}
\right)  =\left(
\begin{array}
[c]{c}%
f_{1}(\lambda_{1},\mu_{1})\\
f_{2}(\lambda_{2},\mu_{2})\\
\vdots\\
f_{n}(\lambda_{n},\mu_{n})
\end{array}
\right)  \text{ or \ }wh=f,\label{14}%
\end{equation}
which is called a \emph{generalized St\"{a}ckel representation. }Multiplying
the HJ equations (\ref{5}) written in the matrix form
\begin{equation}
h=a,\;\;\;\text{with }h=(c,h_{n-1},...,h_{0})^{T}\text{ and \ }a=(c,a_{n-1}%
,...,a_{0})^{T},\label{16}%
\end{equation}
from the left by $w$ one gets $wh=wa$, or according to the condition
(\ref{14}), $f=wa$ i.e.
\begin{equation}
f_{i}(\lambda_{i},\frac{\partial W}{\partial\lambda_{i}})=c\lambda_{i}%
^{n}+a_{n-1}\lambda_{i}^{n-1}+...+a_{0},\;\;\;i=1,\ldots,n\;\label{17}%
\end{equation}
which implies the existence of the separated solution (\ref{6}) for the system
of HJ equations (\ref{5}). In consequence, the system (\ref{5}) splits into a
decoupled set of ODE's
\begin{equation}
f_{i}(\lambda_{i},\frac{dW_{i}}{d\lambda_{i}})=c\lambda_{i}^{n}+a_{n-1}%
\lambda_{i}^{n-1}+...+a_{0},\;\;\;i=1,\ldots,n.\label{17a}%
\end{equation}
which concludes the proof.

Let us now restrict to a special case, when $f_{i}(\lambda_{i},\mu_{i})$ is
quadratic in momenta $\mu_{i}$
\begin{equation}
f_{i}(\lambda_{i},\mu_{i})=f_{i}(\lambda_{i})\mu_{i}^{2}+g_{i}(\lambda
_{i}),\label{17b}%
\end{equation}
(for motivation of this choice see the end of this section) i.e. to the case
when $\mathcal{M}=T^{\ast}Q\times\mathbf{R},$ where $Q$ is some Riemannian
(pseudo-Riemannian) manifold and $T^{\ast}Q$ its cotangent bundle. Then the
condition (\ref{14}) can be put in the form
\begin{equation}
\left(
\begin{array}
[c]{cccc}%
\lambda_{1}^{n-1} & \lambda_{1}^{n-2} & \cdots & 1\\
\lambda_{2}^{n-1} & \lambda_{2}^{n-2} & \cdots & 1\\
\vdots & \vdots & \cdots & \vdots\\
\lambda_{n}^{n-1} & \lambda_{n}^{n-2} & \cdots & 1
\end{array}
\right)  \left(
\begin{array}
[c]{c}%
h_{n-1}\\
h_{n-2}\\
\vdots\\
h_{0}%
\end{array}
\right)  =\left(
\begin{array}
[c]{c}%
f_{1}(\lambda_{1})\mu_{1}^{2}+g_{1}(\lambda_{1})-\lambda_{1}^{n}c\\
f_{2}(\lambda_{2})\mu_{2}^{2}+g_{2}(\lambda_{2})-\lambda_{2}^{n}c\\
\vdots\\
f_{n}(\lambda_{n})\mu_{n}^{2}+g_{n}(\lambda_{n})-\lambda_{n}^{n}c
\end{array}
\right)  .\label{17c}%
\end{equation}
This is a system of linear equations with respect to $h_{i}$. It has the
solution \cite{2}
\begin{equation}
h_{i}=\sum_{k=1}^{n}G_{i}^{kk}(\lambda)\mu_{k}^{2}+V_{i}(\lambda)+c\rho
_{i}(\lambda),\;\;\;i=0,\ldots,n-1\label{30}%
\end{equation}
where
\[
G_{i}^{kk}(\lambda)=-\frac{\partial\rho_{i}}{\partial\lambda_{k}}\frac
{f_{k}(\lambda_{k})}{\Delta_{k}}\text{ \ },\;\;V_{i}(\lambda)=-\sum_{k=1}%
^{n}\frac{\partial\rho_{i}}{\partial\lambda_{k}}\frac{g_{k}(\lambda_{k}%
)}{\Delta_{k}}%
\]
$\Delta_{i}=\prod_{j\neq i}(\lambda_{i}-\lambda_{j})$, and $\rho_{i}(\lambda)$
are the polynomials of order $n$ in $\lambda$, defined by the relation
\begin{equation}
(\lambda-\lambda_{1})(\lambda-\lambda_{2})\ldots(\lambda-\lambda_{n}%
)=\sum_{i=0}^{n}\rho_{i}\lambda^{i}\label{Viete}%
\end{equation}
(i.e. Vi\`{e}te polynomials). It can be shown that in the language of
Riemannian geometry $G_{i}$ are second order contravariant tensors on $Q$; the
tensor $G_{n-1}$ represents a contravariant metric tensor, i.e. the inverse of
a standard covariant metric $g$ on $Q:\,G_{n-1}=g^{-1},$ while the remaining
$n-1$ ones are Killing tensors. From (\ref{17c}) it follows that the constants
of motion for geodesic case have the form
\begin{equation}
\sum_{k=1}^{n}G_{i}^{kk}(\lambda)\mu_{k}^{2}=\sum_{k=1}^{n}(\varphi^{-1}%
)_{i}^{k}\mu_{k}^{2},\label{31}%
\end{equation}
where $\varphi_{i}^{k}=\lambda_{i}^{n-k}/f_{i}(\lambda_{i})$ is a classical
St\"{a}ckel matrix. It is worth mentioning here that the theory of separation
of variables on Riemannian manifolds has been developed by Eisenhart
\cite{eisenhart} who used the concept of Killing tensors to analyze integrals
of geodesic motion. The intrinsic formulation of Eisenhart's theory has been
presented in \cite{benenti} where the concept of Killing web was introduced,
while the application of the theory to bi-Hamiltonian systems has been
presented in \cite{ibort}.

From (\ref{3new}) it can be shown that in $(\lambda,\mu)$ coordinates the
operators $\Pi_{0}$ and$\ \Pi_{1}$ attain the form
\begin{equation}%
\begin{array}
[c]{c}%
\Pi_{0}=\left[
\begin{array}
[c]{c|c}%
\begin{array}
[c]{cc}%
0 & I\\
-I & 0
\end{array}
&
\begin{array}
[c]{c}%
0\\
0
\end{array}
\\\hline
\ast & 0
\end{array}
\right] \\
\\
\Pi_{1}=\left[
\begin{array}
[c]{c|c}%
\begin{array}
[c]{cc}%
0 & \Lambda\\
-\Lambda & 0
\end{array}
&
\begin{array}
[c]{c}%
\partial h_{n-1}/\partial\mu\\
-\partial h_{n-1}/\partial\lambda
\end{array}
\\\hline
\ast & 0
\end{array}
\right]
\end{array}
\label{32}%
\end{equation}
with the diagonal matrix $\Lambda=diag(\lambda_{1,\ldots,}\lambda_{n})$ and
where the symbol $\ast$ denotes the elements that make the matrices
skew-symmetric. In fact there exists a whole family of separated coordinates
$(\lambda^{\prime},\mu^{\prime},c^{\prime})$ which preserve the forms
(\ref{30}) and (\ref{32}) of $h_{i}$\ and $\Pi_{i}$ and which are related to
the set $(\lambda,\mu,c^{\prime})$ by a canonical (gauge) transformation
\begin{equation}
\lambda_{i}^{\prime}=\pm\lambda_{i},\;\;\;\mu_{i}^{\prime}=\pm\mu
_{i}+\vartheta_{i}(\lambda_{i}),\;\;\;i=1,...,n,\label{19}%
\end{equation}
where $\vartheta$ is an arbitrary smooth function. In the separated
coordinates the Poisson pencil $\Pi_{\xi}$ and the chain (\ref{3new}) can be
projected onto symplectic leaf $S_{c^{\prime}}$ of $\Pi_{0}$ ($\dim S=2n $)
producing a non-degenerated Poisson pencil on $S_{c^{\prime}}$ of the form
$\theta_{\xi}=\theta_{1}-\xi\theta_{0},$ where
\begin{equation}
\theta_{0}=\left[
\begin{array}
[c]{cc}%
0 & I\\
-I & 0
\end{array}
\right]  ,\,\,\,\,\,\theta_{1}=\left[
\begin{array}
[c]{cc}%
0 & \Lambda\\
-\Lambda & 0
\end{array}
\right]  ,\label{25}%
\end{equation}
is a nondegenerate Poisson pencil on $S_{c^{\prime}}$. Hence, $S_{c^{\prime}}$
is a Poisson-Nijenhuis manifold where the related Nijenhuis tensor
$\mathcal{N}$ and its adjoint $\mathcal{N}^{\ast}$
\begin{equation}
\mathcal{N}=\theta_{1}\circ\theta_{0}^{-1}=\left[
\begin{array}
[c]{cc}%
\Lambda & 0\\
0 & \Lambda
\end{array}
\right]  \text{ \ \ \ \ \ \ \ \ }\mathcal{N}^{\ast}=\theta_{0}^{-1}\circ
\theta_{1}=\left[
\begin{array}
[c]{cc}%
\Lambda & 0\\
0 & \Lambda
\end{array}
\right] \label{26}%
\end{equation}
are diagonal (here we use the name of Nijenhuis tensor for a second order
tensor with a vanishing Nijenhuis torsion). This motivates the following
definition \cite{morositondo}.

\begin{definition}
The coordinates $(\lambda,\mu,c^{\prime})$ in which $\Pi_{0}$ and $\Pi_{1}$
have the form (\ref{32}) (or, equivalently, in which the tensor $\mathcal{N}%
=\theta_{1}\circ\theta_{0}^{-1}$ has the diagonal form (\ref{26})), are called
the $\emph{Darboux-Nijenhuis}$ (DN) coordinates.
\end{definition}

The tensors $\mathcal{N}$ and $\mathcal{N}^{\ast}$are not equal, since
$\mathcal{N}$ act on the space of vector fields while $\mathcal{N}^{\ast}$
acts on the space of one-forms on $\mathcal{M}$. Notice that $\rho_{i}%
(\lambda)$ are coefficients of minimal polynomial of the Nijenhuis tensor
\begin{equation}
(\det(\lambda I-\mathcal{N}))^{1/2}=\det(\lambda I-\Lambda)=\prod_{i=1}%
^{n}(\lambda-\lambda_{i})=\sum_{i=0}^{n}\rho_{i}\lambda^{i},\;\;\;\rho
_{n}=1.\label{26a}%
\end{equation}
Let us now perform an arbitrary, not necessary canonical, $c^{\prime}%
$-independent, nondegenerate coordinate transformation $(\lambda
,\mu)\rightarrow(q,p)$, preserving a quadratic dependence on momenta in all
$h_{k},\;k=0,\ldots,n-1.$ It can be shown that after such transformation, i.e.
in the variables $(q,p,c^{\prime})$ the operators $\Pi_{0}$ and $\Pi_{1}$ will
have the following form
\begin{equation}%
\begin{array}
[c]{c}%
\Pi_{0}=\left[
\begin{array}
[c]{c|c}%
\begin{array}
[c]{cc}%
0 & -\widetilde{G}(q)\\
\widetilde{G}^{T}(q) & \widetilde{F}(q,p)
\end{array}
&
\begin{array}
[c]{c}%
0\\
0
\end{array}
\\\hline
\ast & 0
\end{array}
\right] \\
\\
\Pi_{1}=\left[  \left.
\begin{array}
[c]{c}%
\begin{array}
[c]{cc}%
0 & G(q)\\
-G^{T}(q) & F(q,p)
\end{array}
\\\hline
\ast
\end{array}
\right|  \left.  \Pi_{0}\circ dh_{n-1}\right.  \right]
\end{array}
\;\label{43a}%
\end{equation}
and the minimal polynomial of the Nijenhuis tensor becomes
\begin{equation}
(\det(\xi I-\mathcal{N}))^{1/2}=\frac{\det(\xi\widetilde{G}+G)}{\det
\widetilde{G}},\label{44}%
\end{equation}
Obviously, in a real situation we start from a given bi-Hamiltonian chain
(\ref{3new}) in arbitrary coordinates $(q,p,c^{\prime}),$ derived by some
method, and find the appropriate transformation to DN coordinates
$(\lambda,\mu,c^{\prime})$. In a typical situation one starts with the
Hamiltonians $h_{i}$ that written in the coordinates $(q,p,c^{\prime})$ are
quadratic in momenta $p$. If the transformation to $(\lambda,\mu,c^{\prime})-$
variables is such that $\lambda$ depends on $q$ only, then the functions
$f_{i}$ must have the form (\ref{17b}). This is precisely the situation one
encounters in the case of bi-cofactor systems. In the next sections we will
apply the ideas of this section in order to separate this class of systems.

\section{Basic facts about bi-cofactor systems}

Let us consider the systems of differential equations (in the flat space
$\mathbf{R}^{n}$) of Newton form:
\begin{equation}
\ddot{q}=M(q)\label{Newton}%
\end{equation}
where $q=(q_{1},\ldots,q_{n})^{T}$, $M(q)=(M_{1}(q),\ldots,M_{n}(q))^{T}$
where $^{T}$ denotes the transpose of a matrix, $q_{i}=q_{i}(t)$, with
$t\in\mathbf{R}$ being an independent variable and where dots denote
differentiation with respect to $t$, so that $\ddot{q}_{i}=d^{2}q_{i}/dt^{2}$
etc. A simple lemma below focuses our attention on those equations of type
(\ref{Newton}) that posses a quadratic in $\overset{.}{q}$ integral of motion.

\begin{lemma}
Let
\[
E(q,\overset{.}{q})=\dot{q}^{T}A(q)\dot{q}+k(q)=\sum_{i,j=1}^{n}\dot{q}%
_{i}A_{ij}(q)\dot{q}_{j}+k(q)
\]
where $A=A^{T}$ is a non-degenerated, $n\times n$ symmetric matrix with
$q$-dependent entries. Then $E$ is an integral of motion for the system
(\ref{Newton}) (that is $\overset{.}{\dot{E}}=0$) if and only if the following
two equations hold:
\begin{equation}
\partial_{i}A_{jk}+\partial_{j}A_{ki}+\partial_{k}A_{ij}=0\,\;\text{for all
\ }i,j,k=1,\ldots,n\label{killing}%
\end{equation}
\begin{equation}
2AM+\nabla k=0\label{quasi}%
\end{equation}
\end{lemma}

Here and in what follows $\partial_{i}=\partial/\partial q_{i}$ and $\nabla
k(q)=(\partial_{1}k,\ldots,\partial_{n}k)$. This lemma can easily be proved by
calculating the derivative $\dot{E}\equiv dE/dt$.

The equation (\ref{killing}) implies that the matrix $A$ is a Killing tensor.
We will restrict ourselves to a class of solutions of (\ref{killing}) that
have the form
\begin{equation}
A=cof(G)\label{A}%
\end{equation}
with
\begin{equation}
G=\alpha qq^{T}+\beta q^{T}+q\beta^{T}+\gamma\label{G}%
\end{equation}
where $cof$ means the cofactor matrix (so that $cof(G)G=\det(G)$ or, in case
when $G$ is invertible, $cof(G)=\det(G)G^{-1}$), $\alpha$ is a real constant,
$\beta=\left(  \beta_{1},\ldots,\beta_{n}\right)  ^{T}$ is a column vector of
constants and where $\gamma$ is a \textit{symmetric} $n\times n$ constant
matrix. It is easy to show that for $n=2$ it is the general solution of
(\ref{killing}). For higher $n$ the general solution of (\ref{killing})
depends on $n(n+1)^{2}(n+2)/12$ parameters (its basis can be found e.g. in
\cite{kalninsmiller}) while (\ref{A}) has only $1+n+n(n+1)/2$ parameters and
is therefore far from being a general solution. It turns out, however, that
this particular solution has interesting properties that make it worth
studying. It originates in a natural way when one considers a broad class of
Poisson pencils of type (\ref{pencil}) (see \cite{stefanjahans}). It leads to
the notion of bi-cofactor systems that admit many differential-algebraic
properties, among them interesting recursion formulas that allow to generate
non-trivial bi-cofactor systems from a trivial (geodesic) flow (see below). It
has been recently generalized to the case of Riemannian manifolds
\cite{crampinsarlet}. In this more general setting it can be demonstrated that
the matrix $G$ given by (\ref{G}) is a conformal Killing tensor, which leads
to the orthogonal separability of geodesic Hamilton-Jacobi equation
\cite{ibort}.

The equation (\ref{quasi}) implies that whenever $\det(A)\neq0$ the force $M$
can be written in the \emph{quasi-potential} form $M=-\frac{1}{2}A^{-1}\nabla
k$, which generalizes the usual potential case and which reduces to the
potential case when $A=\frac{1}{2}I$, where $I$ stands for the identity
matrix. Clearly, in case when our system (\ref{Newton}) has the second -
functionally independent of $E$ - integral of motion of the form
\[
\tilde{E}(q,\overset{.}{q})=\dot{q}^{T}\tilde{A}(q)\dot{q}+\tilde{k}(q)
\]
with an invertible matrix $\tilde{A}$, then it can be written in a
quasi-potential form in two distinct ways. It motivates the following definition:

\begin{definition}
A system of equations
\begin{equation}
\ddot{q}=M(q)=-\frac{1}{2}A^{-1}(q)\nabla k(q)=-\frac{1}{2}\tilde{A}^{-1}%
(q)\nabla\tilde{k}(q)\label{bicofactor}%
\end{equation}
where $A$ and $\tilde{A}$ are two linearly independent matrices of the
cofactor form (\ref{A} )-(\ref{G}) :
\[
A=cof(G),\text{ \ \ }G=\alpha qq^{T}+\beta q^{T}+q\beta^{T}+\gamma
\]
\[
\tilde{A}=cof(\tilde{G}),\text{ \ \ }G=\tilde{\alpha}qq^{T}+\tilde{\beta}%
q^{T}+q\tilde{\beta}^{T}+\tilde{\gamma}%
\]
and where $k=k(q)$ and $\tilde{k}=\tilde{k}(q)$ are two scalar functions, is
called a \emph{bi-cofactor system}.
\end{definition}

The bi-cofactor systems were first studied in \ \cite{PoissonJMP}%
,\cite{stefanjahans} and in \cite{hans}, where they were called cofactor pair
systems. In this article we will deal exactly with this type of systems: we
will show how to perform separation of variables for quasi-potential systems
of the bi-cofactor form.

An important property of such systems is that they actually admit $n$
constants of motion that are quadratic in $\dot{q}$

\begin{theorem}
(Hans Lundmark, \cite{hans}) If the Newton system (\ref{Newton}) has a
bi-cofactor form (\ref{bicofactor}), then it has n integrals of motion of the
form
\begin{equation}
E_{i}(q,\dot{q})=\dot{q}^{T}A_{i}(q)\dot{q}+k_{i}(q),\text{ \ \ }%
i=0,\ldots,n-1\label{calki}%
\end{equation}
where the matrices $A_{i}$ are defined as coefficients in the polynomial
expansion of $cof(G+\xi\tilde{G})$ with respect to the real parameter $\xi$:
\[
cof(G+\xi\tilde{G})=\sum_{i=0}^{n-1}A_{i}\xi^{i}%
\]
with $A_{0}=cof(G)$, $A_{n-1}=cof(\tilde{G})$ and where $k_{0}=k$\ and
$k_{n-1}=\tilde{k}$. Consequently, in case when all matrices $A_{i}$ are
invertible, such system can be written in a quasi-potential form on n distinct
ways:
\begin{equation}
\ddot{q}=M(q)=-\frac{1}{2}A_{i}^{-1}\nabla k_{i},\text{ \ \ }i=0,\ldots
,n-1\label{multi}%
\end{equation}
\end{theorem}

\begin{remark}
\bigskip In the notation as above we have of course $E_{0}=E$ and
$E_{n-1}=\tilde{E}$.
\end{remark}

Of course, for a given pair of matrices $G$ and $\tilde{G}$ not every function
$k$ will have a counterpart $\tilde{k}$ \ that will satisfy the equation
(\ref{bicofactor}). However, there exists a recursion formula that from a
given bi-cofactor system produces a new bi-cofactor system.

\begin{proposition}
\label{rekursjaprop}Let $\ddot{q}=M(q)$ be a bi-cofactor system of the form
(\ref{bicofactor}) with the integrals given by (\ref{calki}). Let also
$k_{\xi}=\sum_{i=0}^{n-1}k_{i}\xi^{i}$ (with $k_{n-1}=\tilde{k}$). Then the
functions $l_{i},$ $i=0,\ldots,n-1$ defined as $l_{\xi}=\sum_{i=0}^{n-1}%
l_{i}\xi^{i}$ through
\begin{equation}
l_{\xi}=\frac{\det(G+\xi\tilde{G})}{\det(\tilde{G})}\tilde{k}-\xi k_{\xi
}\label{rekursja}%
\end{equation}
satisfy the relation $A_{i}^{-1}\nabla l_{i}=A_{j}^{-1}\nabla l_{j}$ for all
$i,j=0,\ldots,n-1$ and are in consequence right hand sides of a new
bi-cofactor system of the form (\ref{multi}) (with $l_{i}$ instead of $k_{i}$).
\end{proposition}

The proof of this statement can be found in \cite{hans}. This formula makes it
possible to produce infinite sequences of bi-cofactor systems starting for
example from a simple geodesic equation $\ddot{q}=0$ which is obviously of a
bi-cofactor form with for example $k_{0}=0$ and $k_{n-1}=1$. Also, it can
easily be inverted in order to express old quasi-potentials $k_{i}$ through
the new quasi-potentials $l_{i}$:
\begin{equation}
k_{\xi}=\frac{1}{\xi}\left(  \frac{\det(G+\xi\tilde{G})}{\det(G)}l-l_{\xi
}\right) \label{antyrekursja}%
\end{equation}
In consequence, by using (\ref{antyrekursja}) it is possible to produce
''lower'' systems from the ''higher'' ones, and if we start from the geodesic
equation $\ddot{q}=0$, we obtain in general the family of ''negative'' systems
different from the sequence of systems obtained from $\ddot{q}=0$ by the use
of (\ref{rekursja}).

It has been proved \cite{stefanjahans}, \cite{hans} that the system
(\ref{bicofactor}) can be embedded in a bi-Hamiltonian system. In order to
make this statement more precise, let us consider a following skew-symmetric
\emph{operator pencil} associated with the system (\ref{bicofactor}):

$\qquad\qquad\qquad\qquad\qquad\qquad\qquad\qquad\qquad\qquad$%
\begin{equation}%
\begin{array}
[c]{c}%
\Pi_{\xi}=\Pi_{1}-\xi\Pi_{0}\equiv\left[
\begin{array}
[c]{c|c}%
\begin{array}
[c]{cc}%
0 & G(q)\\
-G(q) & F(q,p)
\end{array}
&
\begin{array}
[c]{c}%
p\\
M(q)+2cN(q)
\end{array}
\\\hline
\ast & 0
\end{array}
\right] \\
\\
-\xi\left[
\begin{array}
[c]{c|c}%
\begin{array}
[c]{cc}%
0 & -\tilde{G}(q)\\
\tilde{G}(q) & -\tilde{F}(q,p)
\end{array}
&
\begin{array}
[c]{c}%
0\\
-2c\tilde{N}(q)
\end{array}
\\\hline
\ast & 0
\end{array}
\right]
\end{array}
\label{pencil}%
\end{equation}
where both $\Pi_{1}$ and $\Pi_{0}$ are $(2n+1)\times(2n+1)$ matrices/operators
acting in the $(2n+1)$-dimensional space $\mathcal{M}=\mathbf{R}^{2n+1}$ with
Cartesian coordinates labelled with $(q,p,c)$ where $q=\left(  q_{1}%
,\ldots,q_{n}\right)  ^{T},$ $p=\left(  p_{1},\ldots,p_{n}\right)  ^{T},$
$c\in\mathbf{R}$. The $n\times n$ symmetric matrices $G $ and $\ \tilde{G}$
are exactly the matrices that define our system (\ref{bicofactor}). The
$n\times1$ matrices $\ N$ and $\tilde{N}$ are given by
\[
N=\alpha q+\beta,\text{ \ \ \ \ }\tilde{N}=\tilde{\alpha}q+\tilde{\beta}.
\]
The $n\times n$ matrices $F$ and $\tilde{F}$ are defined by
\[
F=Np^{T}-pN^{T}\text{, \ \ \ }\tilde{F}=\tilde{N}p^{T}-p\tilde{N}^{T}.
\]
As usual, the asterisk $\ast$ denotes the elements that makes our matrices
skew-symmetric. This operator pencil is a Poisson pencil precisely due to the
fact, that the term $M(q)$ can be represented as $M(q)=-\frac{1}{2}A_{i}%
^{-1}\nabla k_{i}$ for all $i=0,\ldots n-1$. One should also notice that this
pencil is linear in the variable $c$ and that it has a maximal rank $2n$.
Thus, according to Section 2, the Casimir function $h_{\xi}$ (see Definition
\ref{casimirdef}; obviously, in our variables $d=(\partial/\partial
q,\partial/\partial p,\partial/\partial c)^{T}$ ) of our Poisson operator
$\Pi_{\xi}$ is a polynomial of grad $n$ in $\xi$. In fact, the Casimir
$h_{\xi}$ of $\Pi_{\xi}$ has the form:
\begin{equation}
h_{\xi}(q,p,c)=p^{T}cof(G+\xi\tilde{G})p+k_{\xi}(q)-2c\det(G+\xi\tilde
{G})\label{casimir}%
\end{equation}
where $k_{\xi}(q)=\sum_{i=0}^{n-1}k_{i}(q)\xi^{i}$ are as in Proposition
\ref{rekursjaprop} above. Thus, $h_{\xi}=\sum_{i=0}^{n}h_{i}\xi^{i}$ with the
functions $h_{i}(q,p,c)$ given explicitly by
\begin{equation}%
\begin{array}
[c]{c}%
h_{i}(q,p,c)=E_{i}(q,p)-2cD_{i},\text{ \ \ }i=0,\ldots,n-1\\
\\
h_{n}(q,c)=-2cD_{n}%
\end{array}
\label{hamiltoniany}%
\end{equation}
with $D_{i}=D_{i}(q)$ defined as
\[
\sum_{i=0}^{n}D_{i}(q)\xi^{i}=\det(G+\xi\tilde{G})
\]
so that $D_{0}=\det(G)$ and $D_{n}=\det(\tilde{G})$. The functions $E_{i}$ in
(\ref{hamiltoniany}) are just the constants of motion (\ref{calki}) of our
system (\ref{bicofactor}). By expanding the equation $\Pi_{\xi}\circ dh_{\xi
}=0$ in powers of $\xi$ and comparing the coefficients at equal powers we
obtain the \emph{bi-Hamiltonian chain} \ of the form (\ref{3new}) which by
theorem of Magri \cite{Magri} is completely integrable in the sense of
Liouville. This means that the evolutionary equations
\begin{equation}
\frac{d}{dt}\left[
\begin{array}
[c]{c}%
q\\
p\\
c
\end{array}
\right]  =
\begin{array}
[c]{ccc}%
\Pi_{1}\circ dh_{i} & = & \Pi_{0}\circ dh_{i-1}%
\end{array}
,\text{ }i=1,\ldots,n\label{systemy}%
\end{equation}
associated with (\ref{3new}) are Liouville-integrable.

Let us now investigate the relation of our chain (\ref{3new}) (or of our
Poisson pencil (\ref{pencil})) with the system (\ref{bicofactor}). One can
show, that the last equation in (\ref{systemy}) has at the hyperplane $c=0$
the following form:%

\begin{equation}
\frac{d}{dt}\left[
\begin{array}
[c]{c}%
q\\
p\\
c
\end{array}
\right]  =-2\det(\tilde{G})\left[
\begin{array}
[c]{c}%
p\\
M\\
0
\end{array}
\right] \label{systemdlac=0}%
\end{equation}
so that the hyperplane $c=0$ is invariant with respect to this equation. On
the other hand, if we set $\dot{q}=p$ in (\ref{bicofactor}), we obtain its
equivalent form
\begin{equation}
\frac{d}{dt}\left[
\begin{array}
[c]{c}%
q\\
p
\end{array}
\right]  =\left[
\begin{array}
[c]{c}%
p\\
M
\end{array}
\right] \label{Newtonwqp}%
\end{equation}
which differ from (\ref{systemdlac=0}) only by the coefficient $-2\det
(\tilde{G})$. It means that both systems have the same trajectories in the
$(q,p)$-space, although traversed at different speed.

\begin{lemma}
\label{rescalinglemma}(rescaling) Let $x(t)=(x_{1}(t),\ldots,x_{n}(t))$ be a
solution of a first order differential equation $\dot{x}=X(x)$ in
$\mathbf{R}^{n}$ with $x(0)=x_{0}\in\mathbf{R}^{n}$ and let $\alpha
:\mathbf{R}^{n}\rightarrow\mathbf{R}$ be a scalar function. Then the function
$\ y(t)=x(\tau^{-1}(t))$ with $\tau(t)=\int\frac{1}{\alpha(x(t))}dt$ (chosen
so that $\tau(0)=0$) solves the differential equation $\dot{y}=\alpha(y)X(y)$
with the initial condition $y(0)=x_{0}$.
\end{lemma}

The proof of this lemma is elementary. This lemma implies that knowing a
particular solution of (\ref{systemdlac=0}) we can write down the
corresponding solution of (\ref{Newtonwqp}), just by identifying $\alpha$ with
$-1/(2\det(\tilde{G})).$ In the next section we will show how one can solve
all the equations (\ref{systemy}) (and thus also the bi-cofactor system
(\ref{Newtonwqp})) by a procedure that separates variables in Hamilton-Jacobi
equations that correspond to all Hamiltonians $h_{i}$ of the chain (\ref{3new}).

More information about bi-cofactor systems as well as more detailed
explanations of the facts mentioned above can be found in \cite{stefanjahans}%
,\cite{hans},\cite{oscylator}. In \cite{crampinsarlet} and in
\cite{hansdoktor} one can find a generalization of many of the above mentioned
statements to the case of Riemannian manifolds.

\section{Separation of variables}

In the preceding section we explained how our bi-cofactor system
(\ref{bicofactor}) can be embedded in a bi-Hamiltonian chain (\ref{3new}). The
main goal of this article is to present the procedure that leads to separation
of variables in the Hamilton-Jacobi equations corresponding to the
Hamiltonians $h_{i}$. According to the rescaling lemma above, such a procedure
will yield a solution of our bi-cofactor system as well. This procedure is
based on the results obtained in \cite{m1}, \cite{m2}, \cite{statKdV} and
other papers.

Let us begin by adjusting our coordinate system so that one of the new
coordinates correspond to the foliation of $\mathcal{M}$ into symplectic
leaves of $\Pi_{0}$. We will obtain it by rescaling $c$. Let us thus introduce
the following curvilinear coordinate system in $\mathcal{M}$.
\[
q_{i}^{\prime}=q_{i},\ p_{i}^{\prime}=p_{i},\ i=1,\ldots,n,\ \ \ c^{\prime
}=h_{n}(q,c)=-2c\det(\tilde{G})
\]
(observe that the hypersurfaces $c=0$ and $c^{\prime}=0$ coincide). One should
notice that this transformation does not depend on a particular choice of $k$
and $\tilde{k}$ (since the operator $\Pi_{0}$ does not depend on $k$ and
$\tilde{k})$ but only on the choice of $\tilde{G}$. In what follows we will
write $q_{i}$ and $p_{i}$ instead of $\ q_{i}^{\prime}$ and $p_{i}^{\prime}$.
In the $q,p,c^{\prime}$-variables the chain (\ref{3new}) has the same form;
however, the explicit form of operators $\Pi_{0}$ and $\Pi_{1}$ changes to:
\[
\Pi_{0}=\left[
\begin{array}
[c]{c|c}%
\begin{array}
[c]{cc}%
0 & -\tilde{G}\\
\tilde{G} & -\tilde{F}%
\end{array}
&
\begin{array}
[c]{c}%
0\\
0
\end{array}
\\\hline
\ast & 0
\end{array}
\right]
\]%
\[
\Pi_{1}=\left[
\begin{array}
[c]{c|c}%
\begin{array}
[c]{cc}%
0 & G\\
-G & F
\end{array}
&
\begin{array}
[c]{c}%
-2p\det(\tilde{G})\\
2c^{\prime}(N-G\tilde{G}^{-1}\tilde{N})-2\det(\tilde{G})M
\end{array}
\\\hline
\ast & 0
\end{array}
\right]
\]
(cf (\ref{43a})), while the Hamiltonians $h_{i}$ attain the form:
\[%
\begin{array}
[c]{l}%
h_{i}(q,p,c^{\prime})=E_{i}(q,p)+c^{\prime}\frac{D_{i}}{D_{n}},\text{
\ }i=0,\ldots,n-1\\
h_{n}=c^{\prime}%
\end{array}
\]
In the new variables the symplectic leaves of $\Pi_{0}$ have the desired form
$c^{\prime}=const$. Naturally, the question arises, if we could not choose a
different pencil $\Pi$ so that the formula (\ref{systemdlac=0}) would not
contain the factor $-2\det(\tilde{G})$, which would make Lemma
\ref{rescalinglemma} unnecessary, but we were not able to do so yet.

We are now in position to present the main theorem of this article.

\begin{theorem}
\label{glowny}If the roots $\lambda_{i}=\lambda_{i}(q)$ of the equation
\begin{equation}
\det\left(  G+\lambda\tilde{G}\right)  =0\label{roots}%
\end{equation}
are functionally independent, then in the variables $\lambda=(\lambda
_{1},\ldots\lambda_{n})^{T}$, $\mu=(\mu_{1},\ldots\mu_{n})^{T}$, $c^{\prime}$
given by
\begin{equation}
\lambda_{i}(q)\text{ \ as roots of (\ref{roots})}\label{lambdy}%
\end{equation}
\begin{equation}
\mu_{i}(q,p)=-\frac{1}{2}\frac{\Omega^{T}cof\left(  G+\lambda_{i}(q)\tilde
{G}\right)  p}{\Omega^{T}cof\left(  G+\lambda_{i}(q)\tilde{G}\right)  \Omega
}\text{ \ \ \ \ }i=1,\ldots,n\label{mu}%
\end{equation}
where $\Omega=G\tilde{G}^{-1}\tilde{N}-N$, the operators $\Pi_{0}$ and
$\ \Pi_{1}$ attain the form
\begin{equation}%
\begin{array}
[c]{c}%
\Pi_{0}=\left[
\begin{array}
[c]{c|c}%
\begin{array}
[c]{cc}%
0 & I\\
-I & 0
\end{array}
&
\begin{array}
[c]{c}%
0\\
0
\end{array}
\\\hline
\ast & 0
\end{array}
\right] \\
\\
\Pi_{1}=\left[
\begin{array}
[c]{c|c}%
\begin{array}
[c]{cc}%
0 & \Lambda\\
-\Lambda & 0
\end{array}
&
\begin{array}
[c]{c}%
\partial h_{n-1}/\partial\mu\\
-\partial h_{n-1}/\partial\lambda
\end{array}
\\\hline
\ast & 0
\end{array}
\right]
\end{array}
\label{opwlm}%
\end{equation}
(cf (\ref{43a})), with the diagonal matrix $\Lambda=diag(\lambda_{1,\ldots
,}\lambda_{n})$, while the Hamiltonians $h_{i}$ have the form
\begin{equation}%
\begin{array}
[c]{l}%
h_{i}(\lambda,\mu,c^{\prime})=-\sum_{k=1}^{n}\frac{\partial\rho_{i}}%
{\partial\lambda_{k}}\frac{f_{k}(\lambda_{k},\mu_{k})}{\Delta_{k}}+c^{\prime
}\rho_{i}(\lambda),\text{ \ }i=0,\ldots,n\text{ \ \ \ }(n\geq2)\\
\\
h_{1}(\lambda,\mu,c^{\prime})=f(\lambda,\mu)+c^{\prime}\lambda,\text{ \ }(n=1)
\end{array}
\label{hamwlm}%
\end{equation}
(cf (\ref{30})), with some functions $f_{k}$ depending only on one pair
$\lambda_{k},\mu_{k}$ of the variables, $\Delta_{k}=\prod_{j\neq k}%
(\lambda_{k}-\lambda_{j})$, and where $\rho_{i}(\lambda)$ are the Vi\`{e}te
polynomials (\ref{Viete}). Moreover, in the variables $\lambda,\mu,c^{\prime}%
$the recursion formula (\ref{rekursja}) attains the form
\begin{equation}
l_{\xi}(\lambda)=\det(\xi I-\Lambda)\tilde{k}(\lambda)-\xi k_{\xi}%
(\lambda)\label{rekursjawlm}%
\end{equation}
i.e. the formula (\ref{rekursja}) is invariant with respect to the change of
variables $q\rightarrow\lambda$.
\end{theorem}

\begin{remark}
It is worth mentioning that the coordinates $\lambda(q)$ defined by
(\ref{lambdy}) are in general not orthogonal, and that the gradients
$\nabla\lambda_{i}(q)$ are eigenvectors of the matrix $G\tilde{G}^{-1}$, i.e.
($G+\lambda_{i}\tilde{G})\nabla\lambda_{i}=0$. Moreover, $\nabla\lambda
_{i}(q)$ are $\tilde{G}$-orthogonal: $\nabla\lambda_{i}G\nabla\lambda_{j}=0$
for $i\neq j$. In case when one of the matrices, say $\tilde{G}$, is equal to
the identity matrix (that is when our system (\ref{bicofactor}) becomes
potential) the above transformation (\ref{lambdy})-(\ref{mu}) reduces to the
classical formula for point transformation to separation coordinates for
natural Hamiltonian systems.
\end{remark}

\begin{remark}
The terms $D_{i}/D_{n}$ in the Hamiltonians $h_{i}$ attain the form $\rho
_{i}(\lambda)$ in (\ref{hamwlm}) precisely due to the fact that $D_{i}/D_{n} $
are coefficients in the polynomial expansion of $\det(G+\lambda\tilde{G}%
)/\det(\tilde{G})=\sum_{i=0}^{n}\rho_{i}\lambda^{i}$ \ (cf. (\ref{44} )).
\end{remark}

\begin{remark}
The formulae (\ref{roots})-(\ref{mu}) provide us with a transformation that is
independent of the particular choice of the functions $k$ and $\tilde{k}$ in
the bi-cofactor system (\ref{bicofactor}), i.e. this transformation will
simultaneously separate all the bi-cofactor systems with the same matrices $G
$ and $\tilde{G}.$
\end{remark}

Theorem \ref{glowny} means that the coordinates $(\lambda,\mu,c^{\prime})$ are
Darboux-Nijenhuis coordinates for our operators $\Pi_{0},\Pi_{1}$. According
to results of Section 2, we have

\begin{corollary}
The Hamilton-Jacobi equations for the Hamiltonians $h_{i}(\lambda
,\mu,c^{\prime})$%
\[
h_{i}(\lambda,\frac{\partial W}{\partial\lambda},c^{\prime})=a_{i}\text{
\ \ \ }i=0,1,\ldots,n
\]
where $W(\lambda,a)$ is a generating function for the transformation
$(\lambda,\mu)\rightarrow(b,a)$, separate under the ansatz $W(\lambda
,a)=\sum_{i=1}^{n}W_{i}(\lambda_{i},a)$ into system of ODE's of the form
\begin{equation}
f_{k}\left(  \lambda_{k},\frac{dW_{k}}{d\lambda_{k}}\right)  =c^{\prime
}\lambda_{k}^{n}+a_{1}\lambda_{k}^{n-1}+\cdots+a_{n}\label{ODEs}%
\end{equation}
\end{corollary}

\begin{proof}
The Hamilton-Jacobi equations for Hamiltonians (\ref{hamwlm}) can be treated
as a system of $n$ linear equations for functions $f_{i}$. Applying the Cramer
rule to this system we arrive at (\ref{ODEs}).
\end{proof}

Thus, we are able to find $W$ up to quadratures. If we denote the evolution
parameter associated with $h_{j}$ by $t_{j}$, then in the new variables
$(b,a)$ defined implicitly as usual:
\[
b_{i}=\frac{\partial W(\lambda,a)}{\partial a_{i}}\text{ \ , \ \ }\mu
_{i}=\frac{\partial W(\lambda,a)}{\partial\lambda_{i}}\text{\ \ \ \ \ }%
i=1,\ldots,n
\]
the flow associated with $h_{j}$ has the trivial form (\ref{8}) so that the
transformation $(\lambda,\mu)\rightarrow(b,a)$ simultaneously trivializes
Hamilton equations generated by all the Hamiltonians $h_{i}$.

We will now sketch the proof of Theorem \ref{glowny}. Consider a symplectic
leaf $S_{c^{\prime}}=\{(q,p,c^{\prime}):c^{\prime}=const\}$ of $\Pi_{0}.$ Let
us choose a vector field transversal to the symplectic foliation of $\Pi_{0}$
as $Z=\frac{\partial}{\partial c^{\prime}}.$ It can be shown by direct
calculation that
\[
L_{Z}\Pi_{0}=0,\text{ \ \ \ \ }L_{Z}\Pi_{1}=X\wedge Z\text{ \ \ \ \ with
}X=\Pi_{0}\circ d(Z(h_{n-1}))
\]
where $L_{Z}$ is a Lie derivative operator in the direction of the vector
field $Z$. The above relations guarantee that we can perform a projection of
both $\Pi_{0}$ and $\Pi_{1}$ onto the symplectic leaf $S_{c^{\prime}}$ of
$\Pi_{0}$. The obtained $2n$-dimensional Poisson operators $\theta_{0}$ and
$\theta_{1}$ have the form:%

\[
\theta_{0}(q,p)=\left[
\begin{array}
[c]{cc}%
0 & -\tilde{G}(q)\\
\tilde{G}(q) & -\tilde{F}(q,p)
\end{array}
\right]  \text{ \ \ \ \ \ \ \ \ \ }\theta_{1}(q,p)\equiv\left[
\begin{array}
[c]{cc}%
0 & G(q)\\
-G(q) & F(q,p)
\end{array}
\right]  \text{\ \ }%
\]
The corresponding Nijenhuis tensor $\mathcal{N}=\theta_{1}\circ\theta_{0}%
^{-1}$ has the minimal polynomial of the form $\det(G+\lambda\tilde{G}%
)/\det(\tilde{G})$ \ (cf (\ref{44})) and its roots are precisely the roots of
(\ref{roots}). On the other hand, the roots of the minimal polynomial of
$\mathcal{N}$ define - according to (\ref{26a}) - the first half of the
transformation $(q,p)\rightarrow(\lambda,\mu)$ to DN coordinates in which the
operators $\theta_{0}$ and $\theta_{1}$ have the form (\ref{25}). Due to the
last but one equation in the chain (\ref{3new}) this implies that in the DN
coordinates the operators $\Pi_{0}$ and $\Pi_{1}$ must have the form
(\ref{opwlm}).

We will now show that the remaining part of the transformation
$(q,p)\rightarrow(\lambda,\mu)$ to the DN coordinates, i.e. the expression for
$\mu=\mu(q,p)$, must be of the form (\ref{mu}). We will do it in few steps.

\begin{lemma}
\label{postacX}In the DN coordinates the vector field $X$ has a simple form:
\[
X=\sum_{i=1}^{n}\frac{\partial}{\partial\mu_{i}}%
\]
\end{lemma}

\begin{proof}
\bigskip It is enough to calculate $X=\Pi_{0}\circ d(Z(h_{n-1}))$ in \ the
DN-coordinates:
\[
d(Z(h_{n-1}))=d\left(  \frac{\partial}{\partial c^{\prime}}(h_{n-1})\right)
=d\left(  \frac{D_{n-1}}{D_{n}}\right)  =-d(\lambda_{1}+\cdots+\lambda_{n})
\]
where the last equality is due to the fact that $D_{n-1}/D_{n}$ is precisely
the same term that the last but one in the polynomial expansion of
$\det(G+\lambda\tilde{G})/\det(\tilde{G})$ \ that in turn is precisely the
Vi\`{e}te polynomial $\rho_{n-1}=-(\lambda_{1}+\cdots+\lambda_{n}).$ Thus,
\[
\Pi_{0}\circ d(Z(h_{n-1}))=-\Pi_{0}\circ d(\lambda_{1}+\cdots+\lambda
_{n})=\sum_{i=1}^{n}\frac{\partial}{\partial\mu_{i}}%
\]
\end{proof}

\begin{lemma}
\label{niezalezy}The function
\[
H_{i}(\lambda,\mu,c^{\prime})\equiv\sum_{k=0}^{n}h_{k}(\lambda,\mu,c^{\prime
})\lambda_{i}^{k}%
\]
(i.e. Casimir (\ref{casimir}) written in $(\lambda,\mu,c^{\prime})$ variables
and evaluated at $\lambda_{i}$) depends only on the i-th pair $\lambda_{i}%
,\mu_{i}$ of the variables $\lambda,\mu$ \ i.e.
\[
H_{i}(\lambda,\mu,c^{\prime})=f_{i}(\lambda_{i},\mu_{i})
\]
\ for some function $f_{i}(\lambda_{i},\mu_{i})$.
\end{lemma}

\begin{proof}
We have that $\partial H_{i}/\partial c^{\prime}=$ $\sum_{k=0}^{n}\lambda
_{i}^{k}\partial h_{k}/\partial c^{\prime}=\sum_{k=0}^{n}\lambda_{i}^{k}%
D_{k}/D_{n}=\det(G+\lambda_{i}\tilde{G})/\det(\tilde{G})=0$ due to
(\ref{roots}). For $j\neq i$ we observe that $\partial H_{i}/\partial
\lambda_{j}=\frac{\partial h_{0}}{\partial\lambda_{j}}+\lambda_{i}%
\frac{\partial h_{1}}{\partial\lambda_{j}}+\lambda_{i}^{2}\frac{\partial
h_{2}}{\partial\lambda_{j}}+\ldots+\lambda_{i}^{n}\frac{\partial h_{n}%
}{\partial\lambda_{j}},$ $\ i,j=1,\ldots,n$. On the other hand, due to
(\ref{opwlm}) and (\ref{3new})
\[
-\lambda_{j}\frac{\partial h_{k}}{\partial\lambda_{j}}-\frac{\partial h_{n-1}%
}{\partial\lambda_{j}}\frac{\partial h_{k}}{\partial c^{\prime}}%
=-\frac{\partial h_{k-1}}{\partial\lambda_{j}},\text{ \ \ }k,j=1,\ldots,n
\]
so that
\[
\frac{\partial h_{k}}{\partial\lambda_{j}}=-\frac{\partial h_{n-1}}%
{\partial\lambda_{j}}\left(  \frac{\rho_{k}}{\lambda_{j}}+\frac{\rho_{k-1}%
}{\lambda_{j}^{2}}+\cdots+\frac{\rho_{0}}{\lambda_{j}^{k+1}}\right)
\]
which substituted in the above expression for $\partial H_{i}/\partial
\lambda_{j}$ yields zero. In a similar way one can prove that $\partial
H_{i}/\partial\mu_{j}=0$ for $i\neq j$.
\end{proof}

By Theorem \ref{iff} of Section 2 (and the pages that follow this theorem)
this lemma means that our coordinates $\lambda,\mu$ indeed are separation
coordinates for our systems. We will however continue our line of proof of
Theorem \ref{glowny}.

\begin{corollary}
\bigskip From the above lemma it follows, by Cramer rule, that $h_{i}$ must
have the form (\ref{hamwlm}).
\end{corollary}

\begin{lemma}
\bigskip\cite{statKdV} Let $X=\Pi_{0}d(Z(h_{n-1}))$. Suppose that $X^{r}%
(H_{i})=0$ for some $r=2,3,\ldots\mathbf{.}$and that $X^{k}(H_{i})\neq0$ for
$k=1,\ldots,r-1$. Then
\begin{equation}
\mu_{i}=\frac{X^{r-2}(H_{i})}{X^{r-1}(H_{i})}\label{miwxach}%
\end{equation}
\end{lemma}

In order to prove this lemma it is sufficient to integrate the relation
$X^{r}(H_{i})=0$ twice, using the fact that $X=\sum_{i=1}^{n}\frac{\partial
}{\partial\mu_{i}}$ and Lemma \ref{niezalezy} and use gauge invariance of
DN-coordinates in order to kill integration functions that appear after second integration.

In our case the exponent $r$ that ''kills'' $H_{i}$ is equal to $3$, since
\ in our old coordinates $(q,p,c^{\prime})$ the vector field $X$ has the form
$X=\sum_{i}\left(  \ldots\right)  \frac{\partial}{\partial p_{i}}$ and since
$h_{i}$ are quadratic in $p$. Explicit calculation of expression
(\ref{miwxach}) for $r=3$ and in $q,p,c^{\prime}$-variables yields exactly
(\ref{mu}).

The recursion formula (\ref{rekursjawlm}) is obtained by inserting $I$ and
$\Lambda$ \ as $-\tilde{G}$ and $G$ in (\ref{rekursja}). This concludes the
proof of Theorem \ref{glowny}.

\section{Examples}

We will now illustrate the content of the presented theory with examples. It
is worth to note that the general theory does not provide us with any tools
for calculating the functions $f_{i\text{ }}$ in (\ref{hamwlm}) (or in
(\ref{ODEs})). Instead, we have to calculate these functions each time we
perform the variable separation of a given bi-cofactor system. In case of
bi-cofactor systems however it turns out that the functions $f_{i}$ always
have the form (\ref{17b}).

As a first example, let us consider the family of parabolic separable
potentials introduced in \cite{grammaticos}. They have the form:%

\[
V^{(r)}(q_{1},q_{2})=\sum_{k=0}^{[r/2]}2^{r-2k}\binom{r-k}{k}q_{1}^{2k}%
q_{2}^{r-2k}%
\]
with the other integral of motion given by
\begin{equation}
E^{(r)}=-q_{2}\dot{q}_{1}^{2}+q_{1}\dot{q}_{1}\dot{q}_{2}+q_{1}^{2}%
V^{(r-1)}\label{firstint}%
\end{equation}
We easily find the corresponding matrices $G$ and $\tilde{G}$
\begin{equation}
G=\left[
\begin{array}
[c]{cc}%
0 & -q_{1}/2\\
-q_{1}/2 & -q_{2}%
\end{array}
\right]  \text{ \ },\text{ \ \ }\tilde{G}=\frac{1}{2}I\label{ggex1}%
\end{equation}
$\allowbreak\allowbreak$The recursion formula (\ref{rekursja}), applied to the
geodesic equation $\ddot{q}=0$ with $k_{\xi}=\xi$ (i.e. with $k_{0}=0,$
$k_{1}=1$) produces an infinite family of pairs of quasi-potentials
$k_{0}^{(r)},k_{1}^{(r)},$ $r=0,1,\ldots$ (with $k_{0}^{(0)}=k_{0},k_{1}%
^{(0)}=k_{1}$) such that $(cof(G))^{-1}\nabla k_{0}^{(r)}=cof(\tilde{G}%
))^{-1}\nabla k_{1}^{(r)}\equiv2\nabla k_{1}^{(r)}$. The first few are:%

\begin{equation}%
\begin{array}
[c]{c}%
k_{0}^{(1)}=-q_{1}^{2},\text{ }k_{1}^{(1)}=-2q_{2}\\
k_{0}^{(2)}=2q_{2}q_{1}^{2},\text{ }k_{1}^{(2)}=q_{1}^{2}+4q_{2}^{2}\\
k_{0}^{(3)}=-q_{1}^{4}-4q_{1}^{2}q_{2}^{2},\text{ }k_{1}^{(3)}=-4q_{1}%
^{2}q_{2}-8q_{2}^{3}%
\end{array}
\label{kex1+}%
\end{equation}
where $k_{1}^{(r)}$ correspond to the potentials $V^{(r)}$ up to a factor
$(-1)^{r}$ and where according to (\ref{firstint}) $k_{0}^{(r)}=-q_{1}%
^{2}k_{1}^{(r-1)}$. The corresponding potential-cofactor systems $\ddot
{q}=M^{(r)}(q)=-\frac{1}{2}(cof(G))^{-1}\nabla k_{0}^{(r)}=-\nabla k_{1}%
^{(r)}$ have the form:
\[%
\begin{array}
[c]{c}%
\ddot{q}=M^{(1)}(q)=\left(  0,2\right)  ^{T}\\
\ddot{q}=M^{(2)}(q)=(-2q_{1},-8q_{2})^{T}\\
\ddot{q}=M^{(3)}(q)=\left(  8q_{1}q_{2},4q_{1}^{2}+24q_{2}^{2}\right)  ^{T}%
\end{array}
\]
so that the third one is already non-trivial. Similarly, by applying the
formula (\ref{antyrekursja}), we can produce from the geodesic equation
$\ddot{q}=0$ with $k_{\xi}=1$ (i.e. with $k_{0}^{(0)}=1,$ $k_{1}^{(0)}=0$) the
''negative'' quasi-potentials: The first few of them are of the form
\begin{equation}%
\begin{array}
[c]{c}%
k_{0}^{(-1)}=2q_{2}/q_{1}^{2},\text{ }k_{1}^{(-1)}=-1/q_{1}^{2}\\
\\
k_{0}^{(-2)}=(4q_{2}^{2}+q_{1}^{2})/q_{1}^{4},\text{ }k_{1}^{(-2)}%
=-2q_{2}/q_{1}^{4}\\
\\
k_{0}^{(-3)}=4q_{2}(2q_{2}^{2}+q_{1}^{2})/q_{1}^{6},\text{ }k_{1}%
^{(-3)}=-(4q_{2}^{2}+q_{1}^{2})/q_{1}^{6}%
\end{array}
\label{kex1-}%
\end{equation}
and correspond to potential-cofactor systems $\ddot{q}=M^{(r)}(q)=-\frac{1}%
{2}(cof(G))^{-1}\nabla k_{0}^{(r)}=-\nabla k_{1}^{(r)}$ with
\[%
\begin{array}
[c]{c}%
\ddot{q}=M^{(-1)}(q)=\left(  -2/q_{1}^{3},0\right)  ^{T}\\
\\
\ddot{q}=M^{(-2)}(q)=(-8q_{2}/q_{1}^{5},2/q_{1}^{4})^{T}\\
\\
\ddot{q}=M^{(-3)}(q)=\left(  -4(6q_{2}^{2}+q_{1}^{2})/q_{1}^{7},8q_{2}%
/q_{1}^{6}\right)  ^{T}%
\end{array}
\]
In order to check what variables will separate these systems, we have to solve
the equation (\ref{roots}) with $G$ and $\tilde{G}$ given as in (\ref{ggex1}).
An easy computation yields
\[
\lambda_{1}(q)=q_{2}-\sqrt{q_{1}^{2}+q_{2}^{2}}\ ,\ \lambda_{2}(q)=q_{2}%
+\sqrt{q_{1}^{2}+q_{2}^{2}}%
\]
The formula (\ref{mu}) in this case reads
\[
\mu_{i}=\frac{\sqrt{-\lambda_{1}(q)\lambda_{2}(q)}}{\lambda_{i}(q)}p_{1}%
+p_{2}\text{ , \ \ }i=1,2
\]
and it is immediate to show that the above formulas present the classical
point transformation to the parabolic coordinates. Thus, not only the
potentials $V^{(r)\text{ }}$ but even the corresponding chain (\ref{3new}) is
separable in the parabolic coordinates which is perhaps what we should expect.
After some algebraic manipulations, the above formulas can be inverted to%

\[
q_{1}=\sqrt{-\lambda_{1}\lambda_{2}}\text{ \ \ },\text{ \ \ }q_{2}=\frac{1}%
{2}\left(  \lambda_{1}+\lambda_{2}\right)
\]%

\[
p_{1}=\frac{\sqrt{-\lambda_{1}\lambda_{2}}(\mu_{1}-\mu_{2})}{\lambda
_{1}-\lambda_{2}}\text{ \ \ , \ \ \ }p_{2}=\frac{\mu_{1}\lambda_{1}-\mu
_{2}\lambda_{2}}{\lambda_{1}-\lambda_{2}}%
\]
which makes it possible to express the Hamiltonians $h_{i}$ in the DN
coordinates $(\lambda,\mu,c^{\prime}).$ According to (\ref{hamiltoniany}) and
(\ref{calki}) in the old variables $(q,p,c^{\prime})$ they have the form
\[
h_{i}=\gamma_{i}(q,p)+k_{i}(q)+c^{\prime}\frac{D_{i}}{D_{n}}%
\]
where $\gamma_{i}(q,p)=p^{T}A_{i}(q)p$ is the geodesic part (geodesic
Hamiltonian) of $h_{i}$. Terms $\frac{D_{i}}{D_{n}}=\rho_{i}$ have in the DN
coordinates the form of Vi\`{e}te polynomials (\ref{Viete}) while the form of
$k_{i}$ for a given bi-cofactor system obtained by recursion (\ref{rekursja})
can easily be established either by substituting the above expressions for
$q_{i}(\lambda)$ and \ $p_{i}(\lambda,\mu)$ in (\ref{kex1+})-(\ref{kex1-}) or
from the recursion relation (\ref{rekursjawlm}). The result is
\begin{equation}%
\begin{array}
[c]{c}%
k_{0}^{(-3)}=\dfrac{-(\lambda_{1}+\lambda_{2})(\lambda_{1}^{2}+\lambda_{2}%
^{2})}{\lambda_{1}^{3}\lambda_{2}^{3}}\text{ \ , \ }k_{1}^{(-3)}%
=\dfrac{\lambda_{1}^{2}+\lambda_{1}\lambda_{2}+\lambda_{2}^{2}}{\lambda
_{1}^{3}\lambda_{2}^{3}}\\
\\
k_{0}^{(-2)}=\dfrac{\lambda_{1}^{2}+\lambda_{1}\lambda_{2}+\lambda_{2}^{2}%
}{\lambda_{1}^{2}\lambda_{2}^{2}}\text{ \ , \ }k_{1}^{(-2)}=\dfrac
{-(\lambda_{1}+\lambda_{2})}{\lambda_{1}^{2}\lambda_{2}^{2}}\\
\\
k_{0}^{(-1)}=\dfrac{-(\lambda_{1}+\lambda_{2})}{\lambda_{1}\lambda_{2}}\text{
\ , \ }k_{1}^{(-1)}=\dfrac{-1}{\lambda_{1}\lambda_{2}}\\
\\
k_{0}^{(1)}=\lambda_{1}\lambda_{2}\text{ \ , \ }k_{1}^{(1)}=-(\lambda
_{1}+\lambda_{2})\\
\\
k_{0}^{(2)}=-(\lambda_{1}+\lambda_{2})\lambda_{1}\lambda_{2}\text{ \ ,
\ }k_{1}^{(2)}=\lambda_{1}^{2}+\lambda_{1}\lambda_{2}+\lambda_{2}^{2}\\
\\
k_{0}^{(3)}=\lambda_{1}\lambda_{2}(\lambda_{1}^{2}+\lambda_{1}\lambda
_{2}+\lambda_{2}^{2})\text{\ , \ }k_{1}^{(3)}=-(\lambda_{1}^{3}+\lambda
_{1}^{2}\lambda_{2}+\lambda_{1}\lambda_{2}^{2}+\lambda_{2}^{3})
\end{array}
\label{kawlm}%
\end{equation}
The geodesic Hamiltonians $\gamma_{i}$ have in our case the form
\[
\gamma_{0}=\frac{-\lambda_{2}}{\lambda_{1}-\lambda_{2}}\text{ }\tfrac{1}%
{2}\lambda_{1}\mu_{1}^{2}+\frac{-\lambda_{1}}{\lambda_{2}-\lambda_{1}}\text{
}\tfrac{1}{2}\lambda_{2}\mu_{2}^{2}%
\]%
\[
\gamma_{1}=\frac{1}{\lambda_{1}-\lambda_{2}}\text{ }\tfrac{1}{2}\lambda_{1}%
\mu_{1}^{2}+\frac{1}{\lambda_{2}-\lambda_{1}}\text{ }\tfrac{1}{2}\lambda
_{2}\mu_{2}^{2}%
\]
so that they indeed have the form (\ref{hamwlm}) with $f_{i}$ of the form
(\ref{17b}) (as it has been pointed out at the end of section 2) and we can
identify the functions $f_{i}(\lambda_{i})$ in (\ref{17b}) as $f_{i}%
(\lambda_{i})=\tfrac{1}{2}\lambda_{i}$ which is the form that can be used in
order to solve the inverse Jacobi problem associated with equations
(\ref{ODEs}) and in consequence to separate the system (\ref{bicofactor}).

As a second example we will consider a quite generic (but still
two-dimensional) bi-cofactor system with matrices $G$\ and $\tilde{G}$ of the form%

\[
G=\left[
\begin{array}
[c]{cc}%
q_{1}^{2}+1 & q_{1}q_{2}\\
q_{1}q_{2} & q_{2}^{2}%
\end{array}
\right]  \text{ \ \ \ },\text{ \ \ }\tilde{G}=\left[
\begin{array}
[c]{cc}%
1 & q_{1}\\
q_{1} & 2q_{2}%
\end{array}
\right]
\]
It is no longer potential. The first few of the bi-cofactor systems produced
by the recursion formulae (\ref{rekursja}) and (\ref{antyrekursja}) are
defined by the quasipotentials
\begin{equation}%
\begin{array}
[c]{c}%
k_{0}^{(-2)}=\dfrac{4+2q_{2}+2q_{2}^{2}+q_{1}^{2}}{q_{2}^{2}}\text{ \ ,
\ }k_{1}^{(-2)}=\dfrac{-(2q_{1}^{2}-2q_{2}^{2}+q_{1}^{2}q_{2}-4q_{2})}%
{q_{2}^{3}}\\
\\
k_{0}^{(-1)}=\dfrac{2+q_{2}}{q_{2}}\text{ \ , \ }k_{1}^{(-1)}=\dfrac
{2q_{2}-q_{1}^{2}}{q_{2}^{2}}\\
\\
k_{0}^{(1)}=\dfrac{q_{2}^{2}}{2q_{2}-q_{1}^{2}}\text{ \ , \ }k_{1}%
^{(1)}=\dfrac{q_{2}(2+q_{2})}{2q_{2}-q_{1}^{2}}\\
\\
k_{0}^{(2)}=\dfrac{q_{2}^{3}(2+q_{2})}{(2q_{2}-q_{1}^{2})^{2}}\text{ \ ,
\ }k_{1}^{(2)}=\dfrac{q_{2}^{2}(4+2q_{2}+q_{1}^{2}+q_{2}^{2})}{(2q_{2}%
-q_{1}^{2})^{2}}%
\end{array}
\label{kaex2}%
\end{equation}
and the corresponding forces $M^{(r)}(q)$ are
\[
M^{(-2)}=\dfrac{1}{q_{2}^{4}}\left(  q_{1}(3+q_{2}),q_{2}(4+q_{2})\right)
^{T}%
\]
\[
M^{(-1)}=\dfrac{1}{q_{1}^{3}}\left(  q_{1,}q_{2}\right)  ^{T}%
\]
\[
M^{(1)}=\dfrac{-1}{\left(  q_{1}^{2}-2q_{2}\right)  ^{2}}\left(  q_{1}%
(1+q_{2}),q_{2}^{2}\right)
\]
\[
M^{(2)}=\dfrac{q_{2}}{\left(  q_{1}^{2}-2q_{2}\right)  ^{3}}\left(
q_{1}(q_{1}^{2}+2q_{2}^{2}+4q_{2}+4),q_{2}(q_{1}^{2}+2q_{2}^{2}+2q_{2}%
)\right)  ^{T}%
\]
Also in this case the solutions of (\ref{roots}) can easily be calculated:
\[
\lambda_{1}(q)=\frac{(q_{2}+2+\sqrt{\Delta})q_{2}}{2(q_{1}^{2}-2q_{2})}\text{
\ , \ }\lambda_{2}(q)=\frac{(q_{2}+2-\sqrt{\Delta})q_{2}}{2(q_{1}^{2}-2q_{2})}%
\]
with $\Delta=4q_{1}^{2}+(q_{2}-2)^{2}\geq0$. The coordinate curves given by
these equations consist of the non-confocal ellipses and hyperbolas
\cite{lundmarkprivat} and an arbitrary point $(q_{1},q_{2})$ in the $q$-plane
may lay not only on intersection of an ellipse and a hyperbola (as it was the
case in the classical separability theory) but also on intersection of two
ellipses or two hyperbolas. The formulae (\ref{mu}) for $\mu(q,p)$ are in this
case too complicated to be presented. The inverse relationships are however
still quite compact:
\[
q_{1}=-2\frac{\sqrt{-\lambda_{1}\lambda_{2}(\lambda_{1}+1)(\lambda_{2}+2)}%
}{\lambda_{1}+\lambda_{2}+\lambda_{1}\lambda_{2}}\text{ \ , \ }q_{2}%
=-2\frac{\lambda_{1}\lambda_{2}}{\lambda_{1}+\lambda_{2}+\lambda_{1}%
\lambda_{2}}%
\]
\[
p_{1}=2\frac{\lambda_{1}\lambda_{2}(\mu_{1}\lambda_{1}-\mu_{2}\lambda_{2}%
+\mu_{1}-\mu_{2})}{\lambda_{1}-\lambda_{2}}%
\]
\[
p_{2}=\frac{-(\lambda_{1}+1)(\lambda_{2}+2)\left(  \mu_{1}\left(  \lambda
_{1}^{2}\lambda_{2}-\lambda_{1}^{2}+\lambda_{1}\lambda_{2}\right)  -\mu
_{2}\left(  \lambda_{1}\lambda_{2}^{2}-\lambda_{2}^{2}+\lambda_{1}\lambda
_{2}\right)  \right)  }{(\lambda_{1}-\lambda_{2})\sqrt{-\lambda_{1}\lambda
_{2}(\lambda_{1}+1)(\lambda_{2}+2)}}%
\]
where we have chosen not to simplify the last expression since
\[
-\lambda_{1}\lambda_{2}(\lambda_{1}+1)(\lambda_{2}+2)=q_{1}^{2}q_{2}%
^{2}/(q_{1}^{2}-2q_{2})^{2}%
\]
is always non-negative. Applying these formulas we can express - after long
algebraic manipulations - the geodesic Hamiltonians $\gamma_{i}$ in the DN
coordinates:
\[
\gamma_{0}=\frac{-\lambda_{2}}{\lambda_{1}-\lambda_{2}}\text{ }4\lambda
_{1}^{2}(\lambda_{1}+1)\mu_{1}^{2}+\frac{-\lambda_{1}}{\lambda_{2}-\lambda
_{1}}\text{ }4\lambda_{2}^{2}(\lambda_{2}+1)\mu_{2}^{2}%
\]
\[
\gamma_{1}=\frac{1}{\lambda_{1}-\lambda_{2}}\text{ }4\lambda_{1}^{2}%
(\lambda_{1}+1)\mu_{1}^{2}+\frac{1}{\lambda_{2}-\lambda_{1}}\text{ }%
4\lambda_{2}^{2}(\lambda_{2}+1)\mu_{2}^{2}%
\]
so that $f_{i}(\lambda_{i})=4\lambda_{i}^{2}(\lambda_{i}+1)$ in this case. The
quasi-potentials (\ref{kaex2}) in the DN coordinates must attain the same form
as the quasi-potentials (\ref{kex1+})-(\ref{kex1-}) do, namely \ the form
given by (\ref{kawlm}) since the change of variables (\ref{roots})-(\ref{mu})
is designed so that the recursion (\ref{rekursja}) in the DN coordinates
always attains the form (\ref{rekursjawlm}).

In the end, let us consider a three-dimensional example with matrices $G$ and
$\tilde{G}$ chosen as
\[
G=qq^{T}+\left[
\begin{array}
[c]{lll}%
0 & 0 & {\normalsize 1}\\
0 & 1 & 0\\
1 & 0 & 0
\end{array}
\right]  =\left[
\begin{array}
[c]{lll}%
q_{1}^{2} & q_{1}q_{2} & q_{1}q_{3}+1\\
q_{1}q_{2} & q_{2}^{2}+1 & q_{2}q_{2}\\
q_{1}q_{3}+1 & q_{2}q_{3} & q_{3}^{2}%
\end{array}
\right]
\]
\[
\text{\ }\tilde{G}=\left[
\begin{array}
[c]{l}%
0\\
0\\
1
\end{array}
\right]  q^{T}+q\left[
\begin{array}
[c]{lll}%
0 & 0 & 1
\end{array}
\right]  +\left[
\begin{array}
[c]{lll}%
0 & 0 & 0\\
0 & 1 & 0\\
0 & 0 & 0
\end{array}
\right]  =\left[
\begin{array}
[c]{lll}%
0 & 0 & q_{1}\\
0 & 1 & q_{2}\\
q_{1} & q_{2} & 2q_{3}%
\end{array}
\right]
\]
The recursion formulas (\ref{rekursja}) and (\ref{antyrekursja}) applied to
the geodesic flow $\ddot{q}=0$ yield an infinite sequence of the
quasipotentials $k_{\xi}^{(r)}=\sum_{i=0}^{2}k_{i}^{(r)}\xi^{i}$,
\ $r=\ldots,-1,0,1,\ldots.$Some of them are
\[
k_{0}^{(-1)}=\frac{2q_{1}q_{3}+2q_{1}+1}{2q_{1}q_{3}+q_{2}^{2}+1}\text{ ,
}k_{1}^{(-1)}=\frac{q_{1}(q_{1}+1)}{2q_{1}q_{3}+q_{2}^{2}+1}\text{ , }%
k_{2}^{(-1)}=\frac{q_{1}^{2}}{2q_{1}q_{3}+q_{2}^{2}+1}%
\]
\[
k_{0}^{(1)}=\frac{2q_{1}q_{3}+q_{2}^{2}+1}{q_{1}^{2}}\text{ , }k_{1}%
^{(1)}=\frac{2q_{1}q_{3}+2q_{1}+1}{q_{1}^{2}}\text{ , }k_{2}^{(1)}%
=\frac{2+q_{1}}{q_{1}}%
\]
\[
k_{0}^{(2)}=\frac{(2+q_{1})(2q_{1}q_{3}+q_{2}^{2}+1)}{q_{1}^{3}}\text{ ,
}k_{1}^{(2)}=\frac{2+q_{1}(4q_{3}+4+2q_{1}-q_{2}^{2})}{q_{1}^{3}}\text{ }%
\]
\[
k_{2}^{(2)}=\frac{q_{1}^{2}-2q_{1}q_{3}+2q_{1}+3}{q_{1}^{2}}%
\]
The corresponding forces $M^{(r)}$ are
\[
M^{(-1)}=\frac{-1}{\left(  2q_{1}q_{3}+q_{2}^{2}+1\right)  ^{2}}\left(
q_{1}^{2}\text{ },\text{ }q_{2}(q_{1}+1)\text{ },\text{ }q_{1}q_{3}-1\right)
\]
\[
M^{(1)}=\frac{-1}{q_{1}^{3}}\left(  0,0,1\right)  ^{T}%
\]
\[
M^{(2)}=\frac{-1}{q_{1}^{4}}\left(  q_{1}^{2},q_{1}q_{2},q_{1}q_{3}%
+q_{1}+3\right)  ^{T}%
\]
\[
M^{(3)}=\frac{-1}{q_{1}^{5}}\left(  q_{1}^{2}(q_{1}+4),q_{1}q_{2}%
(q_{1}+3),q_{1}^{2}q_{3}+q_{1}^{2}+3q_{1}+6\right)  ^{T}%
\]
and they fast become complicated with the increasing $|r|$. In this case the
formulas (\ref{roots})-(\ref{mu}) and their inverses are very complicated and
can be handled only with the help of a computer algebra package. We will
therefore quote here only the formulas for $q(\lambda)$
\[
q_{1}=-2\frac{1}{\lambda_{1}+\lambda_{2}+\lambda_{3}+1}%
\]
\[
q_{2}=2\frac{\sqrt{-(\lambda_{1}\lambda_{2}\lambda_{3}+\lambda_{1}\lambda
_{2}+\lambda_{1}\lambda_{3}+\lambda_{2}\lambda_{3}+\lambda_{1}+\lambda
_{2}+\lambda_{3}+1)}}{\lambda_{1}+\lambda_{2}+\lambda_{3}+1}%
\]
\[
q_{3}=\frac{1}{4}\text{ }\frac{\lambda_{1}^{2}+\lambda_{2}^{2}+\lambda_{3}%
^{2}-2(\lambda_{1}\lambda_{2}+\lambda_{1}\lambda_{3}+\lambda_{2}\lambda
_{3}+\lambda_{1}+\lambda_{2}+\lambda_{3})-3}{\lambda_{1}+\lambda_{2}%
+\lambda_{3}+1}%
\]
so that they are built of symmetric polynomials of order three in $\lambda$.
The quasi-potentials $k_{i}^{(r)}$ presented above attain in the above DN
coordinates the form
\[
k_{0}^{(-1)}=\frac{-(\lambda_{1}\lambda_{2}+\lambda_{1}\lambda_{3}+\lambda
_{2}\lambda_{3})}{\lambda_{1}\lambda_{2}\lambda_{3}}\text{ , }k_{1}%
^{(-1)}=\frac{\lambda_{1}+\lambda_{2}+\lambda_{3}}{\lambda_{1}\lambda
_{2}\lambda_{3}}\text{ , }k_{2}^{(-1)}=\frac{-1}{\lambda_{1}\lambda_{2}%
\lambda_{3}}%
\]
\[
k_{0}^{(1)}=-\lambda_{1}\lambda_{2}\lambda_{3}\text{ , }k_{1}^{(1)}%
=\lambda_{1}\lambda_{2}+\lambda_{1}\lambda_{3}+\lambda_{2}\lambda_{3}\text{ ,
}k_{2}^{(1)}=-(\lambda_{1}+\lambda_{2}+\lambda_{3})
\]
\[
k_{0}^{(2)}=(\lambda_{1}\lambda_{2}+\lambda_{1}\lambda_{3}+\lambda_{2}%
\lambda_{3})\lambda_{1}\lambda_{2}\lambda_{3}%
\]
\[
k_{1}^{(2)}=-(2\lambda_{1}\lambda_{2}\lambda_{3}+\lambda_{1}\lambda_{2}%
^{2}+\lambda_{1}\lambda_{3}^{2}+\lambda_{2}\lambda_{1}^{2}+\lambda_{2}%
\lambda_{3}^{2}+\lambda_{3}\lambda_{1}^{2}+\lambda_{3}\lambda_{2}^{2})
\]
\[
k_{2}^{(2)}=\lambda_{1}^{2}+\lambda_{2}^{2}+\lambda_{3}^{2}+\lambda_{1}%
\lambda_{2}+\lambda_{1}\lambda_{3}+\lambda_{2}\lambda_{3}%
\]
which is in accordance with the recursion formula (\ref{rekursjawlm}). The
geodesic Hamiltonians $\gamma_{i}$ have the following structure
\[
\gamma_{0}=\frac{\lambda_{2}\lambda_{3}}{\Delta_{1}}\text{ }4(1+\lambda
_{1})\mu_{1}^{2}+\frac{\lambda_{1}\lambda_{3}}{\Delta_{2}}\text{ }%
4(1+\lambda_{2})\mu_{2}^{2}+\frac{\lambda_{1}\lambda_{2}}{\Delta_{3}}\text{
}4(1+\lambda_{3})\mu_{3}^{2}%
\]
\[
\gamma_{1}=\frac{-(\lambda_{2}+\lambda_{3})}{\Delta_{1}}\text{ }%
4(1+\lambda_{1})\mu_{1}^{2}+\frac{-(\lambda_{1}+\lambda_{3})}{\Delta_{2}%
}\text{ }4(1+\lambda_{2})\mu_{2}^{2}+\frac{-(\lambda_{1}+\lambda_{2})}%
{\Delta_{3}}\text{ }4(1+\lambda_{3})\mu_{3}^{2}%
\]
\[
\gamma_{2}=\frac{1}{\Delta_{1}}\text{ }4(1+\lambda_{1})\mu_{1}^{2}+\frac
{1}{\Delta_{2}}\text{ }4(1+\lambda_{2})\mu_{2}^{2}+\frac{1}{\Delta_{3}}\text{
}4(1+\lambda_{3})\mu_{3}^{2}%
\]
so that they have exactly the form (\ref{hamwlm}) with (\ref{17b}) and with
$f_{i}(\lambda_{i})=4(1+\lambda_{i})$.

\bigskip One can see that in all the above examples the functions
$f_{i}(\lambda_{i})$ do not depend on $i$, i.e. $f_{i}(\lambda_{i}%
)=f(\lambda_{i}).$

\section{Conclusions}

In the present article we performed separation of variables for the recently
discovered class of quasi-potential systems called bi-cofactor systems. These
systems generalize the classical potential systems with additional, quadratic
in momenta, integral of motion in the sense that they reduce to these systems
in case when one of the matrices $G$ and $\tilde{G}$ is the identity matrix.
In this special case the separation formulae (\ref{roots})-(\ref{mu}) reduce
to the well known form the classical separability theory formulas for
separation of natural Hamiltonian systems by a point transformation. In the
general case, however, these formulas do not have the form of a point
transformation and are to our knowledge new in literature. We concluded the
article with some non-trivial examples in which the functions $f_{i}%
(\lambda_{i})$ in (\ref{17b}) actually do not depend on $i$ and we can make a
conjecture that it is always the case in bi-cofactor systems.

\begin{acknowledgement}
We would like to thank S. Rauch-Wojciechowski, H. Lundmark and C. Waksj\"{o}
for interesting discussions during our work on this article. We also thank
referees for useful remarks and references.
\end{acknowledgement}

\end{document}